\documentclass[review,number,sort&compress]{elsarticle}
\biboptions{sort&compress}

\usepackage[utf8]{inputenc}

\usepackage{lineno}
\usepackage{placeins}

\usepackage{graphicx}
\usepackage{amsmath}

\usepackage[babel]{csquotes}

\usepackage[pdftex,unicode]{hyperref}
\usepackage[all]{hypcap}

\usepackage[sort&compress,english]{cleveref}
\crefformat{equation}{#2(#1)#3}
\crefformat{section}{#2section~#1#3}
\Crefformat{section}{#2Section~#1#3}
\crefformat{table}{#2Table~#1#3}
\Crefformat{table}{#2Table~#1#3}
\crefformat{figure}{#2Figure~#1#3}
\Crefformat{figure}{#2Figure~#1#3}

\crefmultiformat{figure}{#2Figures~#1#3}{ and~#2#1#3}{, #2#1#3}{ and~#2#1#3}

\usepackage{todonotes}
\usepackage{siunitx}
\DeclareSIUnit\liter{l}
\DeclareSIUnit[number-unit-product={}]\inch{\SIUnitSymbolArcsecond}

\newcommand{\natf}{$^{24}$Na}
\newcommand{\csots}{$^{137}$Cs}
\newcommand{\naco}{Na$_2$CO$_3$}

\usepackage{epstopdf}

\newcommand{\criticalRate}{\SI{35}{\kilo\hertz}}

\journal{NIM A}

\newcommand*\patchAmsMathEnvironmentForLineno[1]{%
  \expandafter\let\csname old#1\expandafter\endcsname\csname #1\endcsname
  \expandafter\let\csname oldend#1\expandafter\endcsname\csname end#1\endcsname
  \renewenvironment{#1}%
     {\linenomath\csname old#1\endcsname}%
     {\csname oldend#1\endcsname\endlinenomath}}%
\newcommand*\patchBothAmsMathEnvironmentsForLineno[1]{%
  \patchAmsMathEnvironmentForLineno{#1}%
  \patchAmsMathEnvironmentForLineno{#1*}}%
\AtBeginDocument{%
\patchBothAmsMathEnvironmentsForLineno{equation}%
\patchBothAmsMathEnvironmentsForLineno{align}%
\patchBothAmsMathEnvironmentsForLineno{flalign}%
\patchBothAmsMathEnvironmentsForLineno{alignat}%
\patchBothAmsMathEnvironmentsForLineno{gather}%
\patchBothAmsMathEnvironmentsForLineno{multline}%
}

\begin{document}

\begin{frontmatter}
\title{Behavior of a trapezoid-based data acquisition system up to \SI{100}{\kilo\hertz} and beyond}

\author[GU]{S.~Schmidt\corref{cor1}}
	\ead{s1376684@stud.uni-frankfurt.de}

\author[GU]{T.~Heftrich}
\author[GU]{J.~Glorius}
\author[MU]{G.~Hampel}
\author[GSI,GU]{R.~Plag}
\author[GU]{R.~Reifarth}
\author[GU]{Z.~Slavkovská}
\author[GU]{K.~Sonnabend}
\author[MU]{C.~Stieghorst}
\author[MU]{N.~Wiehl}
\author[MU]{S.~Zauner\fnref{WAK}}

\cortext[cor1]{Corresponding author}

\address[GU]{Goethe University, Max-von-Laue-Straße 1, 60438 Frankfurt am Main, Germany}
\address[MU]{Johannes Gutenberg University, Saarstraße 21, 55122 Mainz, Germany}
\address[GSI]{GSI Helmholtzzentrum für Schwerionenforschung GmbH, Planckstrasse 1, 64291 Darmstadt, Germany}
\fntext[WAK]{Current address: WAK Rückbau- und Entsorgungs-GmbH, Eg\-genstein-Leopoldshafen, Germany}

\begin{abstract}
In this work, we investigated the ability of a high-purity germanium detector connected to a trapezoid-filter-based data acquisition system to reliably record signals in spite of high sample activities. By activating multiple \naco{} samples with different Na content, we were able to deduce efficiency, resolution and dead time of the system used as a function of the sample activity. Based on the results, we were able to find a setting which allows measurements of event rates up to \criticalRate{} per readout channel with an energy resolution of \SI{0.3}{\percent} at the \SI{2754}{\kilo\electronvolt} \natf{} line.
\end{abstract}

\end{frontmatter}

\section{Introduction}
Analysis of rare events poses a significant challenge to experimental physics. Commonly, the products from more probable reactions dominate the measured data, therefore requiring high-quality discrimination to extract few events from a large set of input. However, not only data analysis is difficult for these scenarios but also data acquisition. To reduce the time necessary to obtain statistically significant results, luminosity is increased, which causes high count rates from the more probable background reactions. If the dominating reactions bear the same signature as the reactions of interest, effective shielding is impossible, thus requiring the data acquisition system to handle all events. The actual maximum event rate acceptable does not depend solely on the queried type of information but also on the involved detector and readout hardware. Generally, the shorter the time required to process a single event, the higher the maximum usable event rate. However, short processing times usually imply worse resolution properties (e.g. energy resolution).

With the advent of flash-analog-to-digital-converter (flash-ADC)-based readout, the processing time for a single event was decreased significantly: with their extremely short recovery times, flash-ADCs provided dead-time-free streaming of waveform data to a computer, which can do arbitrary pulse analysis. However, such setups were still limited in application because the waveform data either required very fast realtime analysis or large amounts of permanent storage space.

By pairing flash-ADCs with dedicated fast signal analysis processors (field-programmable gate arrays, FPGAs) on a single digitizer board, realtime analysis can be handled by the digitizer and reduces the waveform stream to a few characteristic numbers (e.g. pulse height, integral, and time stamp). For similar count rates as before, this procedure reduces the amount of data written to permanent storage significantly, or, conversely, it enables much higher count rates for similar amounts of storage. 

At the Frankfurt Neutron Source at Stern-Gerlach-Zentrum (FRANZ) \citep{meusel12}, neutron activation analysis with a flux of \SI{1e12}{\per\second\per\square\centi\meter} will allow to study previously inaccessible nuclei, in particular radioactive samples where only small amounts of material are available. Even without the additional activity resulting from the neutron irradiation, the sample activity increases the measured count rate significantly. In order to measure $\gamma$ activities close to \SI{1}{\mega\becquerel} in a $4\pi$ arrangement of two clover detectors with four channels each \citep{dababneh04}, analog data acquisition systems are insufficient as reliable measurements with these systems are limited to a few \si{\kilo\hertz}. Thus, a digitizer-based data acquisition system will be required in order to take full advantage of the high luminosities provided by the neutron source.


Multiple \naco{} samples of different activities were created at the TRIGA (training, research, isotope production, general atomics)-type research reactor in order to characterize the data acquisition system planned for activation experiments at FRANZ.

\section{Sample preparation and setup}
For the samples, a base solution was prepared by dissolving \SI{0.3733}{\gram} of \naco{} powder in \SI{18.5213}{\gram} of distilled water. Based on the molar weight ratio and the density of water at \SI{20}{\degreeCelsius}, the solution was calculated to contain
%
%
%
\SI[round-mode=figures, round-precision=3]{0.008570972266}{\gram} of elementary Na per gram solution. To ensure that the activated volume is constant, this base solution was dissolved in differing amounts of water, yielding the final solutions. The amount extracted (see \cref{tbl:final-solutions}) for the samples was determined by weighing the sample containers before and after filling.
\begin{table*}
	\caption{Sample composition, weight and activity at first measurement. The second column refers to the relative concentration of the base solution (\SI{0.3733}{\gram} \naco{} powder dissolved in \SI{18.5213}{\gram} distilled water) in the solution used for the sample.}
	\label{tbl:final-solutions}
	\centering
	\begin{tabular}{lrrrrr}
		\hline\hline
		sample no. & concentration & sample mass & initial activity \\
		\hline
		1 & \SI{9.45}{\percent} & \SI{0.8558}{\gram} & \SI{40.7}{\kilo\becquerel} \\ 
		2 & \SI{25.10}{\percent} & \SI{0.8506}{\gram} & \SI{108.5}{\kilo\becquerel} \\ 
		3 & \SI{48.86}{\percent} & \SI{0.7866}{\gram} & \SI{193.2}{\kilo\becquerel} \\ 
		4 & \SI{100.00}{\percent} & \SI{0.7443}{\gram} & \SI{374.2}{\kilo\becquerel} \\ 
		\hline\hline
	\end{tabular}
\end{table*}
After neutron-activation, activities beyond \SI{300}{\kilo\becquerel} were expected to enable probing the limits of the detector and data acquisition system.

Using the pneumatic delivery system, each sample was activated for \SI{5}{\minute} in the research reactor. The activated solution was extracted from the activation tube with a pipette and filled into a new (not activated) tube, which was then taken to the detector system. This process introduced a waiting time of approximately \SI{2.3}{\minute}.
The sample was placed at a distance of \SI{6.4}{\centi\meter} from the active volume of a coaxial high-purity germanium detector (Canberra) with an efficiency of \SI{70}{\percent}
%
%
relative to a $\SI{3}{\inch}\times\SI{3}{\inch}$ NaI detector. A \csots{} source was measured simultaneously at a distance of \SI{460.7}{\milli\meter} in order to enable dead-time estimation.

The output from the detector's pre-amplifier (2002CSL with cooled input field-effect transistor) exhibited a signal rise time of $t_\text{rise} = \SI{800}{\nano\second}$ (\SI{10}{\percent} to \SI{90}{\percent}) and a decay time of $t_{1/2,\text{det}} = \SI{33}{\micro\second}$%
%
%
. Without further modification, the output was connected to a CAEN V1724 digitizer, which ran a digital pulse processing firmware for pulse-height analysis (DPP-PHA, version 3.4-128.18) \citep{caen_dpp_pha_application_note}. The digitizer sampled the input voltage at a frequency of \SI{100}{\mega\hertz} in \num{16384} steps of
\SI{0.137}{\milli\volt} each.

\section{Trigger and trapezoid filter algorithm}
In the DPP-PHA firmware, the triggering process is defined by three parameters \citep{tintori13}: the length of an averaging filter $a$, the signal rise time $b$, and the threshold value. Let $\nu(t)$ be the input signal with a discretized time $t$ (i.e. sample counter) and
\begin{align}
	\delta_1(t) &:= \nu_\text{smoothed}(t) - \nu_\text{smoothed}(t - b) \\
	\delta_2(t) &:= \delta_1(t) - \delta_1(t - b)
\intertext{where}
	\nu_\text{smoothed}(t) &:= \frac{1}{a} \, \sum_{i = 0}^{a - 1} \nu(t - i)  
\text{,}
\end{align}
then a trigger is armed if $\delta_2$ exceeds the pre-set threshold value and fired once $\delta_2$ crosses zero.

According to the manufacturer \citep{caen_dpp_pha_application_note}, the trapezoid filter algorithm used is based on previous work \citep{jordanov94}. Basically, the steep voltage rise and slow decay of the input signal can be transformed into a trapezoid of a given length, the plateau height of which is proportional to the original signal's amplitude. The trapezoid filter has six free parameters: the length of the averaging filters for baseline and plateau, the trapezoid rise time corresponding to the classical shaping time, the plateau length, the delay between start of the plateau and start of averaging the plateau amplitude, and the signal decay time constant.

To successfully extract the original pulse height, the amplitude of both the plateau and the baseline are required, so averaging filters allow for the removal of high frequency noise. While averaging over an increasing number of samples generally improves resolution, long smoothing values can become problematic when signals are piled-up. Then, a long smoothing interval might not fit entirely between two consecutive pulses, so data from a previous baseline extrapolation is used for more than a single peak%
%
%
. For higher rates, this can lead to decreased energy resolution. A similar tradeoff between resolution and stability at high rates exists for the trapezoid rise time.

The plateau length should be set to at least fully accommodate the length required for the plateau averaging filter plus the programmable delay from the plateau start to the averaging start. The delay avoids including the unstable trapezoid start in the averaging.

Finally, the decay time is required to successfully cancel out the exponential decay of the input signal and thus maintain a constant trapezoid plateau. This setting corresponds to a classical pole-zero correction.

\smallskip{}

In this experiment, the trigger filter was set to smooth over \num{32}~digital samples (\SI{320}{\nano\second}), while the input rise time was set to \num{31}~digital samples (\SI{310}{\nano\second}). The trigger threshold was set to $\delta_2 \ge 
\SI{2.74}{\milli\volt}$.

For the trapezoid filter, the baseline averaging length was set to \num{4096}~digital samples (\SI{40960}{\nano\second}), while the plateau was averaged over \num{16}~digital samples (\SI{160}{\nano\second}).
The input signal decay time was set to $\tau = \SI{48}{\micro\second}$, corresponding to $t_{1/2,\text{det}} / \ln 2$.
The trapezoid rise time (shaping time), the plateau length and the plateau averaging delay were varied to study their impact on resolution and dead time.

\section{Results}
Using a rise time of $t_\text{rise} = \SI{500}{\nano\second}$, a plateau length of $t_\text{plateau} = \SI{900}{\nano\second}$ and a plateau delay of $t_\text{plateau,delay} = \SI{220}{\nano\second}$, histograms of the different \naco{} samples were obtained in measurement runs of \SI{300}{\second} each (see \cref{fig:full-spectrum}).

\subsection{Peak shape}
\begin{figure*}
	\includegraphics[width=\textwidth]{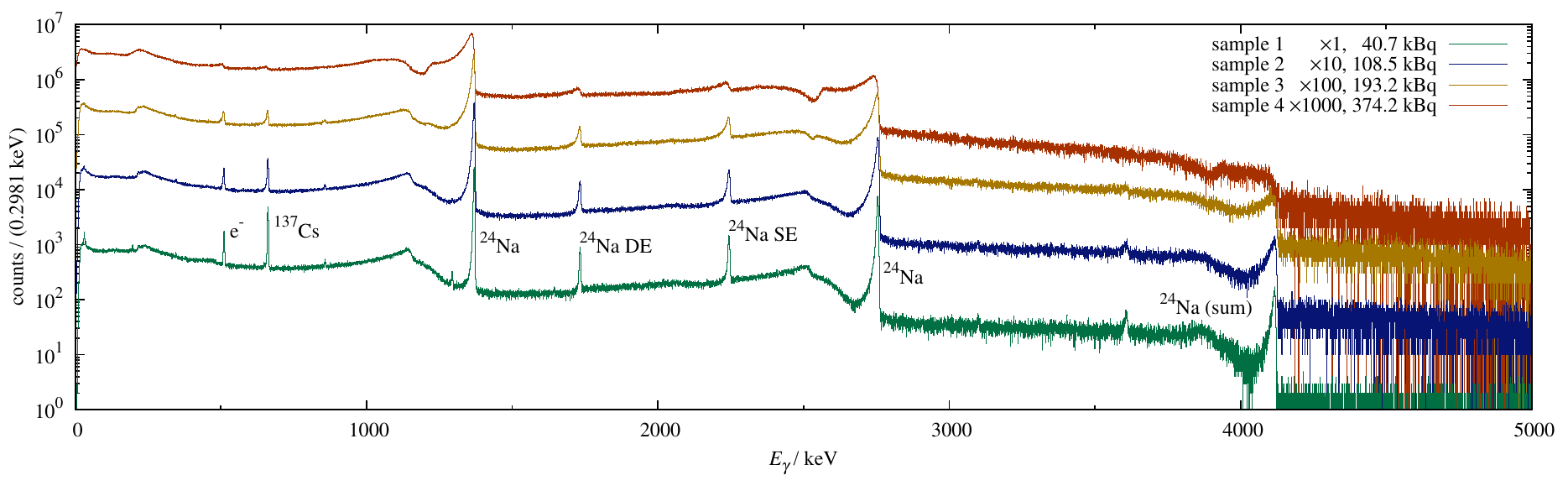}
	\caption{Calibrated pulse-height histogram of the different \naco{} samples from \cref{tbl:final-solutions}. After irradiation, the samples incurred a waiting time of \SI{2.3}{\minute} before data acquisition ran for \SI{5}{\minute}.}
	\label{fig:full-spectrum}
\end{figure*}

\begin{figure*}
	\includegraphics[width=\textwidth]{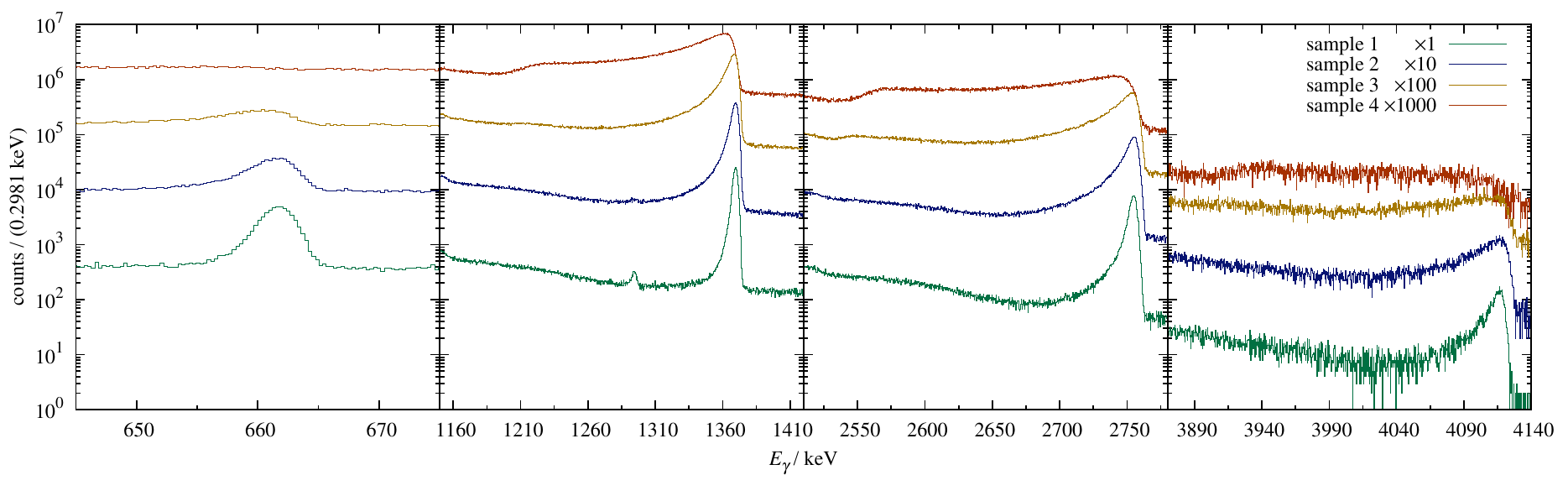}
	\caption{Region around the \csots{} line (\SI{661.657}{\kilo\electronvolt}, \citep{browne07}), the \natf{} lines (\SI{1368.626}{\kilo\electronvolt}, \SI{2754.007}{\kilo\electronvolt}, \citep{firestone07}), and the sum line at \SI{4122.633}{\kilo\electronvolt}, magnified from \cref{fig:full-spectrum}.}
	\label{fig:multipeaks}
\end{figure*}

With increasing \naco{} concentration in the sample, the histogram peaks changed from a gaussian profile to an asymmetric structure with a pronounced secondary peak at lower energies (see the \natf{} lines in \cref{fig:full-spectrum,fig:multipeaks}). While this entire structure contributes to the peak's total area and is thus relevant for activity considerations (see \cref{sec:activity}), assigning a conclusive peak width is difficult: In the case of the \SI{2754}{\kilo\electronvolt} \natf{} line, deviations from a gaussian profile are too large to properly quantify the resolution by a full width at half maximum value. In these cases, the widths are not included in the line width considerations.

\subsection{Activity}\label{sec:activity}
The activity of the the sample with the lowest \naco{} concentration was calculated from the \natf{} sum line: The detected event rate $R_{E_1 + E_2}$ from the sum line of two lines with energies $E_1$ and $E_2$ is
\begin{equation}
	R_{E_1 + E_2} = A ~ \epsilon(E_1) \, I_\gamma(E_1) ~ \epsilon(E_2) \, I_\gamma(E_2) ~\text{,}
\end{equation}
i.e. the source activity $A$ multiplied by the detection probability $\epsilon$ and the absolute gamma intensity, $I_\gamma$. For the lines at energy $E$ (which can be either $E_1$ or $E_2$), the relation is
\begin{equation}
	R_{E_1 + E_2} + R_E = A ~ \epsilon(E_1) \, I_\gamma(E_1)
\end{equation}
where the event rate in the sum line is added as a leading order correction for coincident events. The source activity $A$ can be extracted  from the peak content of the two lines and their sum line:
\begin{equation}
	A = \frac{(R_{E_1 + E_2} + R_{E_1}) \, (R_{E_1 + E_2} + R_{E_2})}{R_{E_1 + E_2}} \label{eqn:real-rate} \text{~.}
\end{equation}

By scaling $A$ with the mass of the \naco{} content, the sample activity for the other samples was calculated (\cref{tbl:final-solutions}).

\subsection{Event rate in the detector, dead time}
\begin{figure}
	\includegraphics[width=\columnwidth]{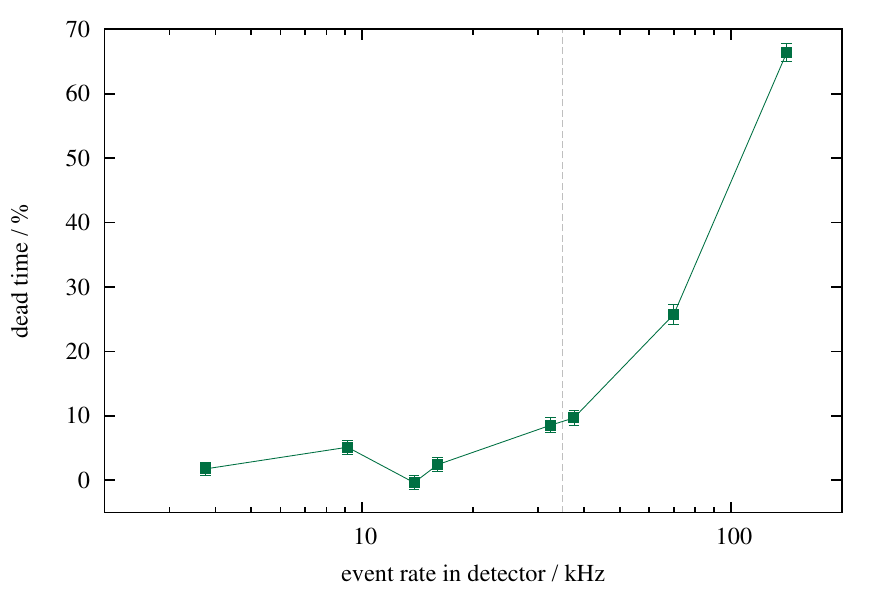}
	\caption{Effect of the \naco{} sample activity on dead time. The dotted gray line marks \criticalRate{} (see \cref{sec:discussion}).}
	\label{fig:activity-vs-deadtime}
\end{figure}

In order to characterize the capabilities of the data acquisition system, the source activity is only indirectly relevant since the distance to the detector influences the event rate strongly. This rate $R_\text{det}$ was computed according to
\begin{equation}
	R_\text{det} = \frac{\sum_{E=0}^{E_\text{max}} N_E / t_\text{m}}{1 - f_\text{dead}} ~\text{,}
\end{equation}
where $N_E$ denotes the number of counts in bin energy bin $E$ and $t_\text{m}$ indicates the measurement duration. $f_\text{dead}$ refers to the dead time, which was determined by comparing the peak content of the \csots{} peak from a measurement where only the \csots{} source was placed in front of the detector (empty run) to the content in a given measurement. The relative transmission $f_\text{trans} = \num{0.852(4)}$ through an empty tube and the sample holder was measured separately. The dead time $f_\text{dead}$ was determined using
\begin{equation}
	f_\text{dead} = 1 - \frac{N_\text{peak} / t_\text{m}}{\left( N_\text{peak,empty} f_\text{trans} \right) / t_\text{m,empty}} \text{~,}
	\label{eqn:dead-time}
\end{equation}
where $N_\text{peak}$ and $t_\text{m}$ denote, respectively, the number of counts in the \csots{} peak and the measurement duration of the corresponding run. The resulting data is depicted in \cref{fig:activity-vs-deadtime}.

\subsection{Efficiency and line width}
\begin{figure}
	\includegraphics[width=\columnwidth]{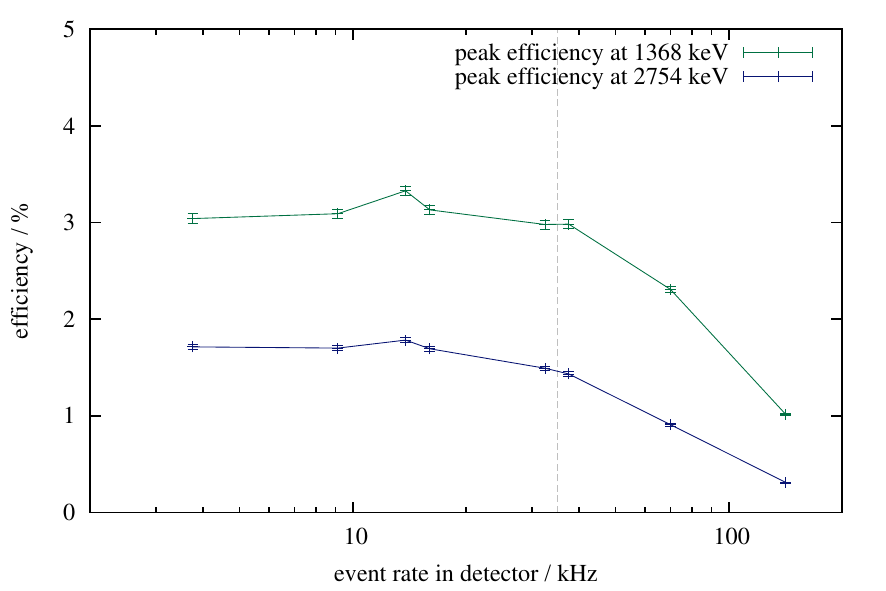}
	\caption{Dependence of the peak efficiency at the position of the \natf{} lines. The dotted gray line marks \criticalRate{} (see \cref{sec:discussion}).}
	\label{fig:activity-vs-efficiencies}
\end{figure}

The peak efficiency $\epsilon(E)$ for the \natf{} lines was derived by comparing the number of events in the line at energy $E$ to the number of decays expected during the measurement $A \, I_\gamma(E) / \lambda \, (1 - \exp(- \lambda \, t_\text{m}))$:
\begin{equation}
	\epsilon(E) = \frac{N_E + N_{E_1 + E_2}}{A \, I_\gamma(E) / \lambda \, (1 - \exp(- \lambda \, t_\text{m}))}
\end{equation}
The dependence of the efficiencies on the event rate are shown in \cref{fig:activity-vs-efficiencies}.

\begin{figure}
	\includegraphics[width=\columnwidth]{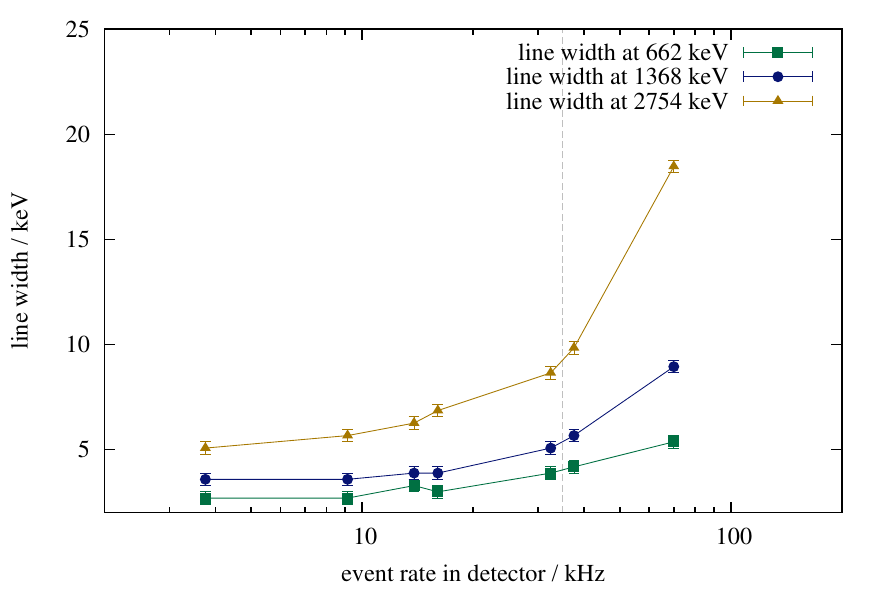}
	\caption{Influence of the activity on measured line width at different energies. The dotted gray line marks \criticalRate{} (see \cref{sec:discussion}).}
	\label{fig:activity-vs-resolution}
\end{figure}

In order to quantify the change in resolution as a function of the rate, the full width at half maximum of the full-energy peak was deduced (see \cref{fig:activity-vs-resolution}).

\subsection{Effect of data acquisition settings}
To determine the influence of acquisition settings on the resulting data, three parameters were varied and the resulting energy resolution and dead time considered. These tests were carried out using sample~3 (\cref{tbl:final-solutions}), so the displayed data cannot refer to the exact same activity. However, the measurements of every single plot were conducted within a timeframe of less than $t = \SI{2.6}{\hour}$. Together with the \natf{} half-life of \SI{14.96}{\hour}, this results in an activity uncertainty of $1 - \exp(-\lambda \, t) \approx \SI{10}{\percent}$\footnote{This error is not included in the error bars.}.

\begin{figure}
	\includegraphics[width=\columnwidth]{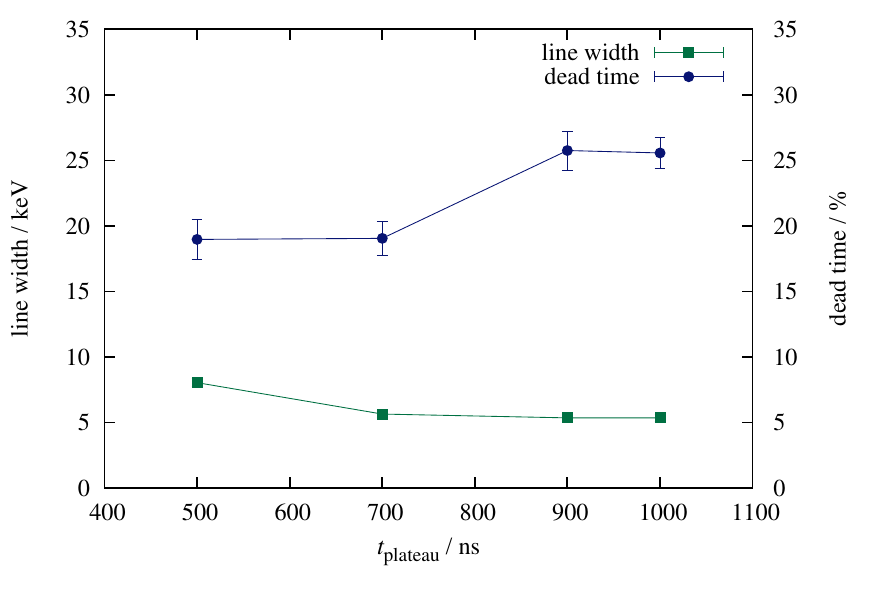}
	\caption{Influence of the plateau length parameter $t_\text{plateau}$ on the dead time and the \csots{} line width (\naco{} sample~3, $t_\text{rise} = \SI{500}{\nano\second}$, $t_\text{plateau,delay} = \SI{220}{\nano\second}$; detector event rates range from \SIrange{56}{70}{\kilo\hertz}).}
	\label{fig:flattop-vs-resolution-deadtime}
\end{figure}

\begin{figure}
	\includegraphics[width=\columnwidth]{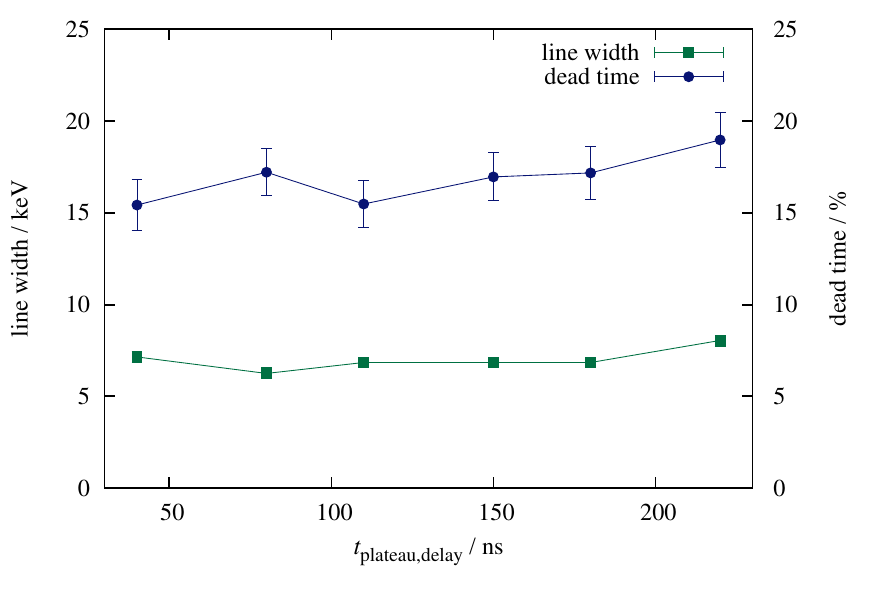}
	\caption{Effect of the plateau averaging delay $t_\text{plateau,delay}$ on the dead time and the \csots{} line width (\naco{} sample~3, $t_\text{rise} = \SI{500}{\nano\second}$, $t_\text{plateau} = \SI{500}{\nano\second}$; detector event rates range from \SIrange{50}{57}{\kilo\hertz}).}
	\label{fig:flattopdelay-vs-resolution-deadtime}
\end{figure}

\begin{figure}
	\includegraphics[width=\columnwidth]{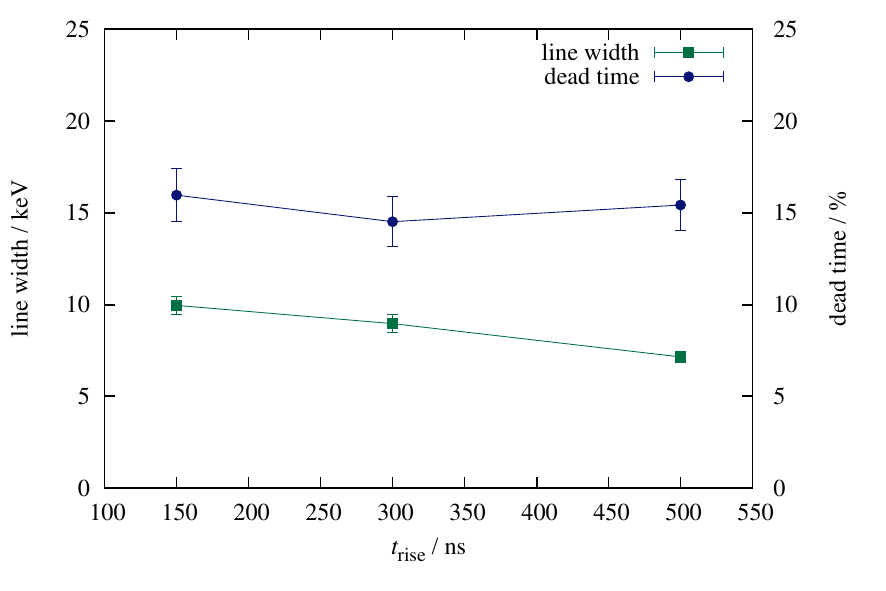}
	\caption{Dependence of the trapezoid rise time $t_\text{rise}$ on the dead time and the \csots{} line width (\naco{} sample~3, $t_\text{plateau} = \SI{500}{\nano\second}$, $t_\text{plateau,delay} = \SI{40}{\nano\second}$; detector event rates range from \SIrange{50}{51}{\kilo\hertz}).}
	\label{fig:shapingtime-vs-resolution-deadtime}
\end{figure}

At first, the length of the trapezoid plateau was changed, testing shorter and longer flat top lengths (\cref{fig:flattop-vs-resolution-deadtime}). The reason for the line width increase at $t_\text{plateau} = \SI{500}{\nano\second}$ is unclear. While further investigations will have to show the reason for this, it might be possible that there is an unstable region at the trapezoid's end. Thus, the end of the peak averaging window (\SI{160}{\nano\second}) would have entered this region because of the comparatively long $t_\text{plateau,delay}$. Based on this possibility, the delay was varied in an attempt to reduce the line width (\cref{fig:flattopdelay-vs-resolution-deadtime}). Finally, with the shortest plateau duration and delay, the effect of the trapezoid rise time $t_\text{rise}$ was probed (\cref{fig:shapingtime-vs-resolution-deadtime}). 

\begin{figure}
	\includegraphics[width=\columnwidth]{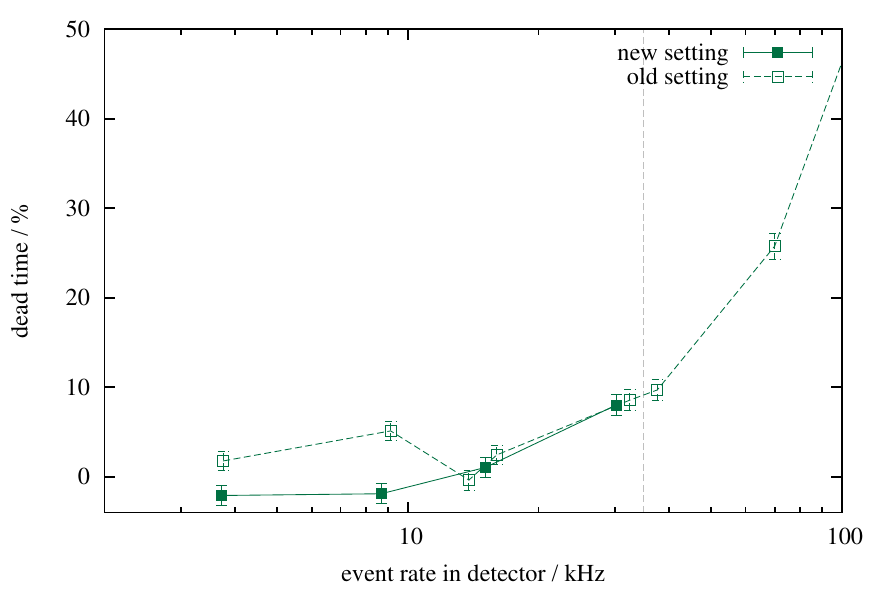}
	\caption{Dead time for the changed digitizer setting (\enquote{new}, $t_\text{rise} = \SI{500}{\nano\second}$, $t_\text{plateau} = \SI{500}{\nano\second}$, $t_\text{plateau,delay} = \SI{80}{\nano\second}$) in comparison to the initial setting (\enquote{old}, $t_\text{rise} = \SI{500}{\nano\second}$, $t_\text{plateau} = \SI{900}{\nano\second}$, $t_\text{plateau,delay} = \SI{220}{\nano\second}$; \cref{fig:activity-vs-resolution}). The dotted gray line marks \criticalRate{} (see \cref{sec:discussion}).}
	\label{fig:activity-vs-deadtime-improved}
\end{figure}

Finally, all \naco{} samples were measured with $t_\text{rise} = \SI{500}{\nano\second}$, $t_\text{plateau} = \SI{500}{\nano\second}$, $t_\text{plateau,delay} = \SI{80}{\nano\second}$ and compared to the initial settings (\cref{fig:activity-vs-deadtime-improved}).

\begin{figure}
	\includegraphics[width=\columnwidth]{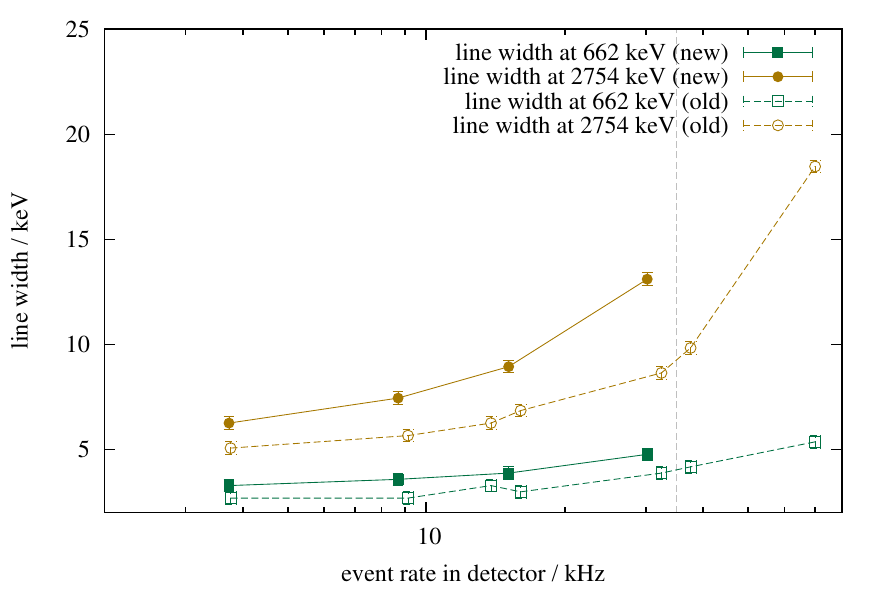}
	\caption{Line width for the changed digitizer setting (\enquote{new}, $t_\text{rise} = \SI{500}{\nano\second}$, $t_\text{plateau} = \SI{500}{\nano\second}$, $t_\text{plateau,delay} = \SI{80}{\nano\second}$) in comparison to the initial setting (\enquote{old}, $t_\text{rise} = \SI{500}{\nano\second}$, $t_\text{plateau} = \SI{900}{\nano\second}$, $t_\text{plateau,delay} = \SI{220}{\nano\second}$; \cref{fig:activity-vs-resolution}). The line width at \SI{1368}{\kilo\electronvolt} follows the same trend as the \SI{662}{\kilo\electronvolt} and the \SI{2754}{\kilo\electronvolt} line. 
	The dotted gray line marks \criticalRate{} (see \cref{sec:discussion}).}
	\label{fig:activity-vs-resolution-improved}
\end{figure}

\section{Discussion}\label{sec:discussion}
In general, the activity-dependent behavior of efficiency, dead time and line width follow the expected trend: With increasing activity, the efficiency decreases while line width and dead time increase. However, all of these three quantities show a change in their slope around \criticalRate{}, indicating that the digitizer-based data acquisition system is well-suited to handle these event rates in a single channel.

Trying to find a possibly better digitizer setting, $t_\text{plateau,delay}$ was found to allow smaller line widths at a setting around \SI{80}{\nano\second} (\cref{fig:flattopdelay-vs-resolution-deadtime}). Increasing the $t_\text{plateau}$ parameter showed increased dead times, but also provided higher resolution (\cref{fig:flattop-vs-resolution-deadtime}). Unlike the other two parameters, the $t_\text{rise}$ parameter (\cref{fig:shapingtime-vs-resolution-deadtime}) does not show a clear trend; an increased number of data points would help finding the optimum setting for this parameter. However, the continuously changing activities complicate the attempt to find similar activities in a limited sample set.

The comparison between the updated and the initial settings is not affected by this difficulty. It shows that both settings provide similarly low dead times in the range up to \criticalRate{} (\cref{fig:activity-vs-deadtime-improved}), indicating that the data acquisition system can maintain its stability at high rates even when using non-optimal settings. However, the increase in line width and thus the decrease in resolution (\cref{fig:activity-vs-resolution-improved}) shows that the \enquote{new} setting was inferior to the \enquote{old} one: Around \SI{30}{\kilo\hertz}, the \csots{} line widths differ by approximately \SI{21}{\percent}, the width of the \SI{2754}{\kilo\electronvolt} line even increases from \SI{9}{\kilo\electronvolt} to \SI{14}{\kilo\electronvolt} by about \SI{50}{\percent}.

Since resolution and stability at high rates limit each other mutually, one would expect that the \enquote{new} setting is able to support higher decay rates with lower dead times. However, \cref{fig:activity-vs-deadtime-improved} shows no clear improvement, so it is possible that the dead time is not dominated by the digitizer-based readout but rather by the pre-amplifier of the detector system. This is supported by experience gained from measuring the freshly-irradiated sample~4: a scope connected to the detector's pre-amplifier displayed alternately heavily piled-up detector signals and a zero line. It is unclear if such effects have worsened the results obtained for lower event rates. They might possibly be the cause for the change in slope in the efficiency, dead time and line width plots (\cref{fig:activity-vs-efficiencies,fig:activity-vs-deadtime,fig:activity-vs-resolution,fig:activity-vs-deadtime-improved,fig:activity-vs-resolution-improved}).

\section{Conclusion}
The data acquisition studied in this work was found to support event rates of \criticalRate{} in a single channel. For higher rates, the dead time increases beyond \SI{10}{\percent}, which can be unacceptable for certain experiments.

Approaching this rate, a setting was found which provides line widths of \SI{9}{\kilo\electronvolt} at \SI{2754}{\kilo\electronvolt} corresponding to \SI{0.3}{\percent}.

The findings in this work imply that a count rate of about \SI{300}{\kilo\hertz} can be measured by connecting this data acquisition system to an eight-channel segmented detector. However, issues with the detector's pre-amplifier may have caused additional dead time, which limited the acceptable count rate. In principle it therefore may be possible that a different amplifier type could allow the $4\pi$ arrangement of two clover detectors to measure $\gamma$ activities close to \SI{1}{\mega\becquerel}.

\section{Acknowledgements}
This project was supported by the GIF Research Grant No. \hbox{G -1051-103.7/2009} and the HGF Young Investigator Project VH-NG-327.

\FloatBarrier
\bibliographystyle{model1a-num-names}
\bibliography{references}

\end{document}